\documentstyle[12pt]{article}
\date{}
\def\Poincare{Poincar\'e }
\newcommand{\sect}[1]{\addtocounter{section}{1}\begin{center}
        \normalsize{\bf\arabic{section}.~ #1}\end{center}\vspace*{4mm}}

\newcommand{\bra}[1]{\langle #1 |}
\newcommand{\ket}[1]{| #1 \rangle}

\def\<{\langle}
\def\>{\rangle}
\def\A{{\cal A}}

\def\M{{\cal M}}
\def\H{{\cal H}}
\def\I{{\cal I}}

\def\R{{\cal R}}
\def\P{{\cal P}}

\def\beq{\begin{equation}}
\def\eeq{\end{equation}}
\def\beqarray{\begin{eqnarray}}
\def\eeqarray{\end{eqnarray}}
\def\and{\; ,\qquad\mbox{and}\qquad}
\def\half{{\textstyle{1\over2}}}
\def\thalf{{\textstyle{3\over2}}}
\def\nhalf{{\textstyle{9\over2}}}
\def\turd{{\textstyle{1\over3}}}

\def\fturd{{\textstyle{5\over3}}}
\def\fourth{{\textstyle{1\over4}}}

\def\sxth{{\textstyle{1\over 6}}}

\def\one{\mbox{1}}
\begin{document}

\title { \Poincare Covariant Quark Models of  \\
Baryon Form Factors}

\author{F. Coester}
\maketitle

\centerline{\it Physics Division, Argonne National Laboratory,
Argonne, IL 60439-4843, USA}

\vspace{0.5cm}

\centerline{and}

\vspace{0.5cm}

\centerline{\large D.O. Riska}

\vspace{0.5cm}

\centerline{\it Department of Physics, 00014 University of Helsinki,
Finland}

\vspace{1cm}

\centerline{\bf Abstract}
\vspace{0.5cm}

\Poincare covariant quark models of the
the nucleon, the $\Delta$ resonance and their excitations are
explored. The baryon states are represented by 
eigenfunctions of the four-velocity and a
confining mass operator, which reproduces the empirical 
spectrum up to $\sim$ 1700 MeV to an accuracy 
of $\sim 6\%$. Models of constituent quark
currents provide the relations
between ground-state properties and transition amplitudes.
\vspace*{20mm}

hep-ph/9707388
\newpage
\small
\sect{ Introduction}
\vspace{0.5cm}
The nonrelativistic quark model involves two underlying assumptions,
neither one of which is required by its phenomenological success: 
(1) Constituent
quarks have all the material properties of free particles,
which happen to be confined by a potential. 
(2) The dynamics of the quarks is nonrelativistic. 
In the case of constituent quarks the relation of
typical hadron sizes to the constituent quark masses is such that the
Galilean approximation can provide no more than order of magnitude
estimates. This limitation does not, however, affect applications to hadron
spectroscopy, as such applications involve only the little group $SU(2)$.

The  requirement of \Poincare covariance \cite{Haa} of the states 
and the current operators is  essential for the 
consideration of form factors and transition amplitudes,
and can be met in the context
of relativistic particle dynamics \cite{Kei}.
The success of quark models in accounting for hadron spectra does not depend 
on the assumption that constituent quark is a particle system. 
While states are represented by functions of three position, spin, flavor and
color variables no additional particle assumptions are required.
The principal features needed to account for empirical mass spectra, 
are the symmetry properties of the mass operator \cite{Bijk}.
For this no quark mass is required \cite{Fu,Rav1}. 
Quark-mass parameters may however appear in the 
current-operator model as masses 
relate momenta to velocities.
Here we explore a phenomenology, which does not 
assume a particle structure of hadronic quark currents. 
The 
principal purpose
is to establish a basis for 
empirical relations between the 
electromagnetic structure of nucleons and  baryon resonances. 
The conjecture is
that this does not require any detailed assumptions about the material 
properties of constituent quark  beside
the fundamental symmetries.

In order to facilitate comparison with models based on the assumption of 
a constituent quark-particle structure we note that
the particle structure determines
free-quark currents, which satisfy all symmetry requirements in the
absence of quark interactions.
Confinement is implemented in this framework
by a  modification of the free-particle mass
operator with appropriate symmetry and spectral features.
The choice of a ``form of dynamics'' \cite{Dir}
involves the choice of a ``kinematic subgroup'' for which
the free and interacting unitary representations are identical.
 Modifications of the 
free-particle mass operator  always destroy
the \Poincare covariance of 
free-particle currents, which is restored by appropriate interaction
currents.

In the ``instant-form'' of dynamics the 
center-of-mass position (the Newton-Wigner operator) 
remains a kinematic quantity. In this form
all  boosts are affected by the dynamics, which implies that in an impulse
approximation the momenta of the contributing constituents cannot be related 
kinematically. This form of dynamics is appropriate when all
relevant boosts are approximately Galilean.
The ``light front form'' is unique in that it allows
a consistent formulation of an impulse approximation, which retains
the main qualitative features of the nonrelativistic impulse 
approximation for space-like 
momentum transfer.
In it initial and
final states are related by kinematic Lorentz transformations and 
the momenta of contributing constituents are related kinematically. 
In the impulse approximation
a kinematic three-momentum transfer to the target is taken up by a
single constituent.  Calculations of form factors based on this
approach  \cite{BeTe}-\cite{Cardb} have, however, not established that 
a quark-particle structure is either required or ruled out.
In the ``point form'' of dynamics the full Lorentz group 
is the kinematic subgroup and all four translations  depend on the dynamics.
The point form
has the advantage that Lorentz covariance is readily
implemented by the operator structure 
and that translation covariance may easily be imposed on the 
matrix elements because the four components of the four-momentum
commute. 
 Only in 
this form  is there a kinematic transformation to relative four-momenta and  
Lorentz covariant spinor
wave functions.
In general the four-momentum operator is specified by the mass operator 
and three kinematic variables. The choice of these variables determines the
{\it form of kinematics}. In the point form the kinematic variables 
are the three independent components of the total four-velocity. 

For a description of confined quark systems without quark 
particle structure one may
start with a unitary  \Poincare representation where the 
mass operator is a multiple of the identity. 
On the Hilbert space of states
so defined quark dynamics is introduced by modification 
the mass operator.
The harmonic oscillator model \cite{Fu,Rav1,Rav2} provides a
convenient prototype. The kinematics obtained in this manner is in the 
point-form. Quark  currents  
may be specified  by appropriate
velocities related to the internal  quark momenta.
Once the eigenfunctions of the mass operator are known
the unitary transformations, which relates different forms of 
kinematics are readily 
available.
The symmetry requirements by themselves leave considerable freedom in the 
construction of models. Single-quark 
currents constructed with instant-form
kinematics automatically include features that are interaction currents 
with point-form kinematics.

 Here we introduce  a simple confining mass operator 
which fits
the empirical
mass spectrum to an accuracy of 6\% or better,  up to $\sim 1700$ MeV.
We use this spectroscopic model to formulate an exploratory approach to 
a phenomenology of quark currents. The simplest quark current
is a function of
spin and flavor, which depends on quark momenta only through the 
spectator constraints and hence the form of kinematics. 
This oversimplified  model shares many well-known 
features with nonrelativistic quark models, 
while respecting all requirements
of \Poincare 
covariance. Well-known gross features of the nucleon form factors
determine the values and $Q^2$ dependence of all transition amplitudes.
The model can be refined  to include 
explicit dependence on orbital quark velocities.

Ultimately the question arises of the relation to quantum field 
theory.
Point-form Hamiltonian dynamics may easily be related to constraint 
dynamics, in which
all \Poincare transformations are implemented kinematically and
states are represented by equivalence classes of  functions with a 
semidefinite inner product. Covariant constraint dynamics 
\cite{Jall} provides the bridge
to Bethe-Salpeter formalisms \cite{Tjon} 
and models, which attempt to implement features of 
quantum field theory \cite{Rob}. Euclidean Green functions,
which satisfy reflection positivity \cite {Ost} provide
a basis for unitary representations of the \Poincare
group. These issues are, however, beyond the scope of 
the present paper.

This paper is divided into into 7 sections. In section 2 we define 
the Hilbert
space of 3-quark wave functions. In section 3 the 
mass operator and its spectrum are described. General 
properties of current operators and current matrices are reviewed
 in section 4. The model for the kernels of the single-quark currents
is constructed in section 5 to produce a rough description of 
the nucleon properties.
The model so constructed is applied to inelastic 
transitions in section 6. Section 7 contains a summary.

\sect{The Hilbert Space of 3-Quark States }

The states in the baryon spectrum
are described by  vectors in the little Hilbert space $\H_\ell$
\cite{Haa}, which is 
the representation 
space of the direct product of
the little group, $SU(2)$, with flavor  and color $SU(3)$.
Concretely these states are
realized by functions, 
$\phi$,
of three quark positions $\vec r_i$, three spin variables $\mu_i$
and three flavor variables $f_i$, 
which are symmetric under permutations
and invariant under translations $\vec r_i\to \vec r_i+\vec a$. 
The translational invariance is realized by
expressing the wave functions  in terms of 
Jacobi coordinates
\beqarray
&&\vec r:={1\over \sqrt{2}}(\vec r_1-\vec r_2),\cr\cr &&
\vec \rho:=\sqrt{{2\over 3}}(\vec r_3-{\vec r_1+\vec r_2\over 2}).
\eeqarray
Representations of the full \Poincare group
obtain on the 
tensor product, $\H:=\H_\ell\otimes \H_c$, of the little Hilbert space
$\H_\ell$ with the Hilbert space $\H_c$ 
of functions of the four-velocity $v$, which is 
specified by 3 independent components. Translation are generated
the four-momentum operator $P=\M v$. 
Any confining self-adjoint
mass operator $\M$, independent of $v$,  satisfies  all 
relativistic symmetry requirements if it is invariant under rotations. 
At the level of spectroscopy
alone there is no difference
between relativistic and 
nonrelativistic quark models because the Galilean rest energy
operator
satisfies all the symmetry requirements of  a mass operator.

The \Poincare invariant inner
product of the functions  representing baryon states is defined as
\beq
(\Psi,\Psi)=\int d^4v 2
\delta(v^2+1)\,\theta(v^0)\int d^3\kappa\int
d^3k|\Psi(v,\vec \kappa,\vec k)|^2,
\label{SPR}
\eeq
where $\vec \kappa$ 
and $\vec k$ are  the momenta conjugate to $\vec\rho$ and $\vec r$ .
Summation over spin and flavor variables is implied.
Under a Lorentz transformations $v\rightarrow \Lambda v$ the vectors
$\vec\kappa$ and $\vec k$ and the three quark spin variables $\mu_1,\mu_2$, 
$\mu_3$ undergo Wigner rotations $R_W(\Lambda,v)$,
\beq
R_W(\Lambda,v):= B^{-1}(\Lambda v)\Lambda B(v)\; .
\eeq

 Note that with 
canonical boosts 
\beq
R_W(\Lambda_v,v):= B^{-1}(\Lambda_v v)\Lambda B(v)=\one
\eeq
for any rotationless Lorentz transformation 
$\Lambda_v$ in the direction of $\vec v$.

For  heuristic constructions of impulse 
currents it is convenient
to define internal four-momenta $p$ and $q$ by
\beq
p:=B(v)\{0,\vec \kappa\}\and q:=B(v) \{0,\vec k\}\; ,
\label{BPQ}
\eeq
so that
$p^2=|\vec \kappa|^2$  and $q^2=|\vec k|^2$ .

The construction of unitary representations of the \Poincare group   
sketched here naturally leads
to point-form kinematics. Once the eigenfunctions of the mass operator 
are known it is easy
to realize unitary transformations to other forms of kinematics
explicitly.
Let $\Psi_n(v,\vec \kappa,\vec k)$ be 
eigenfunctions of $\M$, with eigenvalues 
 $M_n$.
Any state  $\Psi=\sum_n \Psi_n c_n$ can be represented by functions 
$\Psi(\vec P,\vec p, \vec q)$
normalized as
\beq
(\Psi,\Psi)=\int d^3 P\int d^3 p\int d^3q |
\Psi(\vec P, \vec p,\vec q)|^2\; .
\eeq
The unitary transformation 
$\Psi(v,\vec \kappa, \vec k)\to \Psi(\vec P,\vec p, \vec q)$ 
is specified by the 
variable transformation  $\{v,\vec \kappa, \vec k, n\} 
\to \{\vec P,\vec p, \vec q,n\}$
where $\vec p =\vec p(v,\vec \kappa)$ and $\vec q 
=\vec q(v,\vec k)$ are specified by
eq. (\ref {BPQ})
and $ \vec P = M_n\vec v$ in each term of the sum over $n$.

\newpage

\sect{The Mass Operator and its Eigenfunctions }
\setcounter{equation}{0}

As suggested by the
empirical baryon spectrum we shall choose the square of the zero order
mass operator to be 
\beq
{\cal M}_0^2=3[\vec \kappa\,^2+\vec k\,^2+\omega^4(\vec\rho\,^2+\vec
r\,^2)]\; ,
\label{MM}
\eeq
where $\omega$ is a phenomenological parameter. 
Note that this mass operator does not contain
a quark mass.
The mass operator (\ref{MM})
commutes with the  velocity $v$ and the spin operator
$\vec j$, and is independent of $v$. 
It is therefore \Poincare invariant as required, and completely
symmetric under permutations of the coordinates. 

To the zero-order mass operator $\M_0$ we
add a ``hyperfine'' correction,
$\M'$ of the form
\beq
{\cal M}^{'}=-C\sum_{i<j}\,
\vec \lambda^i\cdot\vec \lambda^j\;\vec \sigma^i\cdot
\vec \sigma^j\; , 
\label{HYP}
\eeq
where $C$ is a constant. Here $\vec \lambda^i$ is the $SU(3)$ flavor
generator of the $i$th quark, and $\vec \sigma^i$ is the
corresponding spin matrix. 

 The
flavor-spin structure of the hyperfine interaction  (\ref{HYP}) 
corresponds to the spin-flavor part of the interaction
mediated by the exchange of the octet of light pseudoscalar
bosons, which are the Goldstone bosons of the
approximate chiral symmetry of QCD \cite {GlRi}. Here
the main rationale for it is that leads to a very satisfactory
description of the observed baryon spectrum up to $\sim 1700$ MeV
as shown below.
If all of the hyperfine interaction 
between the constituent quarks were assumed to arise mainly
from exchange of a
pseudoscalar  boson octet \cite {GlRi}, the constant $C$
should be replaced by a radial function with vanishing volume integral,
and which changes sign and takes
the form of a pion exchange Yukawa function at large distances.

Taking $C$ to be a constant implies an unrealistically long range for
the hyperfine correction, and does not allow for the large empirical
splitting of the $SD$-shell in the baryon spectrum. This deficiency may
remedied, without loss of integrability, by adding an angular-momentum
dependent correction of the following form to the mass operator:
\beq
{\M}''=A[(\vec r \times \vec k)^2+(\vec \rho\times \vec
\kappa )^2]\;,
\eeq
where $A$ is a constant. The inclusion of such
a term makes it possible to extend the satisfactory description of the
baryon spectra beyond 1700 MeV.

The eigenfunctions of the mass operator 
\beq
{\cal M}={\cal M}_0+{\cal M}^{'}+{\cal M}^{''},
\label{MHF}
\eeq
are 
linear combinations of functions of the form
\beq
\Psi_{N,L,[X]}(v, \vec \kappa,  
\vec k,f_1,f_2,f_3,\sigma_1,\sigma_2,\sigma_3)
=
\phi_{N,L,[X]}(v, \vec \kappa, \vec k)\,
\Phi_{S,T,[X]}(f_1,f_2,f_3,\sigma_1,\sigma_2,\sigma_3)
\; .
\label{WFN}
\eeq
Here $[X]=[3],[21],[111]$ are Young patterns, which label 
the symmetric, mixed and antisymmetric irreducible 
representations of the permutation group $S_3$ for the orbital wave function.
The variables  $\sigma_i$ and $f_i$ 
$(i=1,2,3)$ are quark spin and quark flavor
labels. As we restrict the treatment here to the
nucleon and the $\Delta$ spectra the flavor indices $f_i$ 
correspond to the usual isospin indices. 

The functions $\phi_{N,L,[X]}( \vec \kappa, \vec k)$ 
are  products of harmonic oscillator 
functions $\varphi_{nlm}$ of $\vec\kappa$ and $\vec k$, which  are
completely determined by the orbital
angular momentum $L$,
the principal quantum number $N\geq L$,  and the symmetry character $[X]$.
The explicit wave functions $\Psi$ for the baryon states in Table 1 are
listed in Table 2.

In the unitarily equivalent instant-form representation
one has
\beq
\Psi_{N,L,[X],S,T}(\vec P, \vec p,  
\vec q,f_1,f_2,f_3,\sigma_1,\sigma_2,\sigma_3)
=
\phi_{N,L,[X],S,T}(\vec P, \vec p, \vec q)\,
\Phi_{S,T,[X]}(f_1,f_2,f_3,\sigma_1,\sigma_2,\sigma_3)
\; .
\label{WFNI}
\eeq
In this form the orbital functions depend on $S$ and $T$.

The
eigenvalues of the  mass operator $\M=\M_0+\M'+\M''$
are 
\beq
\epsilon=\sqrt{6(N+3)}\: \omega+K(C,A)\; ,
\eeq
where $K$ is a hyperfine correction, which is listed for the states
with $N\leq 2,\,\, L\leq 1$ up to $\sim$ 1700 MeV in Table 1. 

The  mass operator, $\M$, contains no quark mass parameter. 
The parameters 
$\omega$ and $C$  may be determined by the nucleon mass
and the real part of the $\Delta(1232)$ pole position:
$
M(N)=\sqrt{18}\:\omega-14C=939\,{\rm MeV},
$
and 
$
M(\Delta(1232)=\sqrt{18}\omega-4C
=1211\, {\rm MeV}.
$
These equations yield the 
values $\omega$ = 311 MeV and $C$ = 27.1 MeV. The parameter $A$
in the correction term (3.3) is determined by  the empirical
difference of  the $SD$ shell resonances $N(1440)$ and $N(1720)$ to be
$A=43$ MeV. The calculated
resonance energies (averaged over the multiplets) obtained with these
parameter values are listed in Table 1. The values so calculated 
deviate from the empirical values by about $\sim 6\%$ at most. 

\sect{ Current Operators }
\setcounter{equation}{0}

The quark current density operators $I^\mu(x)$ have to satisfy
the following covariance conditions
\beq
U^\dagger(\Lambda)I^\mu(x)
U(\Lambda)=\Lambda^\mu\, _\nu I^\nu(\Lambda^{-1}x)\; ,
\eeq
for arbitrary Lorentz transformations $\Lambda$. In the case
of space-time translations this requirement takes the form
\beq
e^{iP\cdot a}I^\mu(x)e^{-P\cdot a}=I^\mu(x+a)\; .
\eeq
Current
conservation requires that
\beq
[P_\nu,I^\nu(0)]=0\; .
\label{CON}
\eeq
The
current density operators are assumed to be 
operator valued tempered 
distributions. Under this assumption
their Fourier transforms exist and are also
operator valued tempered distributions: 
\beq
\tilde{I}(Q):= {1\over (2\pi)^4}\int d^4xe^{-iQ\cdot
x}I^\mu(x)\; . 
\eeq
The covariance relations
\beq
U^\dagger(\Lambda)\tilde{I}^\mu(Q)U(\Lambda)=\Lambda^\mu\,_\nu
\tilde{I}^\nu(\Lambda^{-1}Q)\; ,
\eeq
and
\beq
[P^\nu,\tilde{I}^\mu(Q)]=Q^\nu\tilde{I}^\mu(Q)\; ,
\label{MCON}
\eeq
follow from these requirements.

Let $\ket{p,j,\sigma,\tau,\zeta}\equiv \ket{M,v,j,\sigma,\tau,\zeta}$  be
eigenstates of the four-momentum operator $P=\M v$ and the canonical spin,
with $\sigma $ an 
eigenvalue of $j_z$ and $\zeta=\pm 1$ the intrinsic parity.
It then follows from
the translation covariance (\ref{MCON}) that
the matrix elements of $\tilde I^\mu(Q)$ and
$I^\mu(0)$ are related by
\beqarray
&&\bra{\kappa',\tau',\sigma',j',v',M'}
\tilde{I}^\mu(Q)\ket{M,v,j,\sigma,\tau,\kappa}
\cr\cr &&=\delta^{(4)}(M'v'-Mv-Q)
\bra{\kappa',\tau',\sigma',j',v',M'} 
I^\mu(0)\ket{M,v,j,\sigma,\tau,\kappa }\; .
\label{MCUR}
\eeqarray

It will be convenient  to
define a time-like  unit vectors $u$, orthogonal to $v'-v$ by
\beq
u:= {v'+v\over {\sqrt{-(v'+v)^2}}}= {v'+v\over 2 v^0}\; ,
\eeq
where
\beq
v^0:= -u\cdot v = - u\cdot v'={\sqrt{1+\eta}}\; ,\qquad \eta:= {v^0}^2-1
 \; .
\eeq
We may assume, without loss of generality, that the plane defined by
$v'$ and $v$ is the $(t,z)$-plane.

The subgroup, $O(2)$, that leaves $v$ , $v'$  invariant consists of the
rotations $\R_z(\varphi)$ about the z-axis and the reflection $\P_y$
of the y-axis.
Under these transformations the 
charge longitudinal components, $ I^0(0)$ and
 $ I_z(0)$  are scalars and the transverse
current $ I_{\pm 1}:=\half[(  I_x(0)\pm iI_y(0)]$ 
transforms as an $O(2)$ vector:
\beqarray
&&U^\dagger[\R_z(\varphi)]  I^0(0)U[R_z(\varphi)] = I^0(0)\; ,
\qquad U^\dagger(\P_y)  I^0(0)U(P_y) = I^0(0)\;,\cr\cr &&
U^\dagger[\R_z(\varphi)] I_z(0)U[R_z(\varphi)] = I_z(0)\; ,
\qquad U^\dagger(\P_y)  I_z(0)U(P_y) = I_z(0)\;,\cr\cr &&
U^\dagger[\R_z(\varphi)]  I_{\pm 1}(0)U[R_z(\varphi)]=
e^{\pm i\varphi}  I_{\pm}(0)
\; , \qquad 
U^\dagger(\P_y)  I_{\pm 1}(0)U(\P_y)= I_{\mp 1}(0)\; .
\cr &&
\eeqarray
States transform according to the rules
\beqarray
&&U^\dagger[\R_z(\varphi)]\ket{M,v,j,\sigma,\tau,\zeta} = 
\ket{M,v,j,\sigma,\tau,\zeta} e^{i\sigma\varphi}\; ,\cr\cr &&
 U^\dagger(\P_y) \ket{M,v,j,\sigma,\tau,\zeta}
 = \ket{M,v,j,-\sigma,\tau, \zeta}\,\zeta\, (-1)^{j-\sigma}\; .
\eeqarray
It follows that the matrix elements of the scalar operators $ I^0(0)$ and
$I_z(0)$  vanish for $\sigma'\not= \sigma$ and  the matrix 
elements of  $ I_{\pm 1}(0)$ vanish unless $\sigma'-\sigma=\pm 1$.
Current conservation implies that
$
P\cdot I(0)=I(0)\cdot P\; .
$

The irreducible representations of $O(2)$  are labeled by  $|\sigma|$.
The use of the Wigner-Eckart theorem for $O(2)$ is straightforward.
The $O(2)$ covariant matrix elements are equal to invariant reduced 
matrix elements (form factors)
multiplied by $O(2)$ Clebsch-Gordan coefficients,
$C^{|\sigma'|,k,|\sigma|}_{\sigma',\beta,\sigma}$,  
\newpage
\beqarray
&&\<\zeta ',\sigma,j',v',M'|u\cdot I(0)|M,v,j,\sigma,\zeta\>=
C^{|\sigma'|,0,|\sigma|}_{\sigma,0,\sigma}\,
(\zeta ',j',M'\Vert I_0(v'\cdot v,|\sigma|)\Vert M,j,\zeta)\; ,\cr\cr &&
\<\zeta ',\sigma,j',v'.M'|\half (v'-v)\cdot I(0)|M,v,j,\sigma,\zeta\>=
C^{|\sigma'|,0,|\sigma|}_{\sigma,0,\sigma}\,
(\zeta ',j',M'\Vert I_z(v'\cdot v ,|\sigma|)\Vert M,j,\zeta)\sqrt{\eta}\; ,
\cr\cr &&
\<\zeta ',\sigma'
,j',v', M'|I_\beta (0)|M,v ,j,\sigma, \zeta\>
=C^{|\sigma'|,1,|\sigma|}_{\sigma',\beta ,\sigma}\,
(\zeta ',j',M'\Vert I_1(v'\cdot v ,|\sigma| ) \Vert M,j,\zeta)\; .
\eeqarray
where $\beta=\pm1$.
The non-vanishing Clebsch-Gordan coefficients 
are equal to $\pm 1$. We may choose
\beq
C^{|\sigma'|,k,|\sigma|}_{\sigma',\beta,\sigma}= 1  \qquad 
\mbox{for}\quad \sigma\geq 0\; .
\eeq
The full current operator is determined by these reduced matrix elements.

The definition of the momentum transfer $Q:=M'v'-Mv$
implies that
\beq
{Q^2+(M-M')^2\over 4MM'}\equiv 
{\tilde  Q^2\over (M'+M)^2}\;=\; \eta \;=v_z^2={v_z'}^2
=\fourth(v'-v)^2\;,
\eeq
where
\beq
\tilde Q:=Q+(Q\cdot u)\,u= \half(M'+M)\,(v'-v)\; .
\eeq
It is customary to define Lorentz invariant ``charge'' and ``longitudinal'' 
current components, $I_{CH}(0)$ and $I_L(0)$, so that
\beq
I(0)=I_{CH}(0)v' 
+I_L(0)\left[{v'-v\over 2 \sqrt{\eta}}\,\sqrt{1+\eta}+\sqrt{\eta}u\right]
+I_\perp(0)\; .
\eeq

The transitions between the nucleons and the excited
states are described by helicity amplitudes $A_\lambda(Q^2)$ 
($\lambda=1/2, 3/2$), 
defined as matrix elements of the
transverse current multiplied by a conventional invariant 
factor:
\beq
A_\lambda(Q^2):=  \sqrt{4 \pi \alpha \over 2E_\gamma}
\bra{\zeta ',\lambda,j',v', M'}I_1(0)
\ket {M,v ,j,\lambda-1, \zeta}\; ,
\label{HEL}
\eeq
where $\alpha$ is the fine structure constant and 
\beq
E_\gamma:= -v'\cdot Q +{Q^2\over 4M'M}= {{M'}^2-M^2\over 2M'}
\eeq
is the photon energy for radiative decay.
\newpage
\sect{Current Kernels and Nucleon Form factors}
\setcounter{equation}{0}
Current density operators $\tilde I^\mu(Q)$ can be represented by 
kernels  $(\vec k\,',\vec \kappa\,',v'|\I^\mu(Q)|v,\vec \kappa,\vec k)$ 
such that matrix elements
$\bra{f}I^\mu(0)\ket{a}$ are related to the wave function
(\ref{WFN}) by 
\beqarray
\bra{f} I^\mu(0)\ket{i}&=& \int d^4 Q\delta^{4}(M'v'-Mv-Q)\cr\cr &\times &
\int d^3 k' \int d^3\kappa '
\int d^3\kappa \int d^3 k\Psi^*_f(v',\vec \kappa\, ',\vec k\, ')
(\vec k\,',\vec \kappa\,',v'|\I^\mu(Q)|v,\vec \kappa,\vec k)
\Psi_i(v,\vec \kappa,\vec k)\; .\cr &&
\eeqarray
For the purpose of specifying impulse 
currents we define three formal
quark-momentum transfers
\beqarray
 Q_1&:=&m(Q,v',v)(v'-v)
-{1\over 2}{\sqrt{2\over 3}}(\tilde p'-\tilde p)+\sqrt{{1\over
2}} (\tilde q'-\tilde q) \; ,
\cr\cr
 Q_2&:=& m(Q,v',v)(v'-v)-{1\over 2}
\sqrt{{2\over 3}}(\tilde p'-\tilde p)-\sqrt{{1\over
2}}(\tilde q'-\tilde q)\; ,
\cr\cr
 Q_3&:=& m(Q,v',v)(v'-v) + \sqrt{{2\over 3}} (\tilde p'-\tilde p)\; ,
\label{MOM}
\eeqarray
where $\tilde p:=p+(u\cdot p)u$ and $\tilde q:= q+(u\cdot q)u$. 
The scale
factor $m(Q,v',v)$ introduced here plays the role of an effective quark
mass. It should be emphasized that there is great latitude in the
choice of this function. 
We  specify the impulse current, $I^\mu_i$, by momentum constraints
$Q_k=0, \forall k\not=i$.

A requirement that the impulse 
constraints be kinematic
significantly limits possible choices depending on the form of kinematics.
With point-form kinematics 
the scale factor is restricted to functions  $m(\eta)$.
This leads to definite relations 
between nucleon elastic form factors and 
transition amplitudes which we will explore below. 
With $m(Q,v',v)=\sxth 
\sqrt{\tilde Q^2/\eta}$ the impulse constraint (\ref{MOM})
is kinematic with instant-form kinematics.

Because of the complete antisymmetry of the baryon wave functions it is
sufficient to consider the current matrix elements of only one
constituent, e.g. $i=3$. It follows from the constraints 
$v\cdot p=0$ and $v'\cdot p'=0$  that
\beq
p_\perp=\kappa_\perp =p'_\perp=\kappa'_\perp\;,\qquad 
p^0=p_zv_z/v^0\; ,\qquad {p'}^0=p'_zv'_z/v^0\; , \qquad p_z=v^0 \kappa_z\; .
\eeq
 It follows from the constraints 
$v\cdot p=0$ and $v'\cdot p'=0$  that
\beq
p_\perp=\kappa_\perp =p'_\perp=\kappa'_\perp\;,\qquad 
p^0=p_zv_z/v^0\; ,\qquad {p'}^0=p'_zv'_z/v^0\; , \qquad p_z=v^0 \kappa_z\; .
\eeq
The momentum constraints  $Q_1=Q_2=0$ imply that $Q_3= 3m(v'-v)$,
and
\beq
(\vec k\,',\vec \kappa\,',v'|\I^\mu|v, \vec \kappa, \vec k,)
=3\I^\mu_3(v',v)
\delta[ \vec \kappa\,'- \vec \kappa
-{\sqrt{6}}\,{m\over v^0}(\vec v\,'-\vec v)]
\delta(\vec k\,'- \vec k)\; .
\eeq
with $\mu=\{0,\perp\}$. 

A very lean model for the current
kernels is the following:
\newpage
\beqarray
\I_3^0(v',v)\equiv u\cdot\I_3(v',v)&=&
\left[\half
\lambda_3^{(3)}f_3+
\half{\sqrt{\turd}}\lambda^{(3)}_8f_8\right]\; \otimes \one\; ,
\cr\cr
\vec \I_3(v',v)&=&
i\thalf [\vec \sigma_3\times (\vec v\,'-\vec v)]\,
\left[\half
\lambda_3^{(3)}g_3+
\half{\sqrt{\turd}}\lambda^{(3)}_8g_8 \right]\otimes \one
\; .
\label{LEAN}
\eeqarray
With point-form kinematics it follows from
\beq
\half(|\vec \kappa|^2+|\vec \kappa\,'|^2) 
=\fourth|\vec 
\kappa'+\vec \kappa|^2 + {6m^2(\eta) \eta\over (1+\eta)}\; ,
\eeq
that
all matrix elements of impulse currents are proportional to 
a function $F_0(\eta)$,
which for $\eta<< 1$ can be approximated by the usual 
dipole form factor:
\beq
F_0(\eta):=
\exp\left(-{6m^2(\eta)\eta\over \omega^2( 1+\eta)}\right)
\approx \left({1\over 1+(3m^2(0)\eta/ \omega^2)}\right)^2\; ,
\eeq
 with $m^2(0)=\fturd \omega^2$. The dipole form obtains 
for all values of $\eta$ with the choice
\beq
m^2(\eta)= \omega^2{1+\eta\over 3\eta}\,ln(1+5 \eta)\; .
\eeq

The isoscalar and isovector nucleon magnetic moments are respectively
\beq
\mu_{IS} =  \mu_p+\mu_n=g_8, \quad \mu_{IV}=  \mu_p-\mu_n=5g_3\; .
\eeq
Agreement with the corresponding empirical values 
are obtained with $g_8=.88$ and $g_3=.94$ 

For the magnetic moment of the $\Delta^{++}$ and the 
$N\rightarrow p$ transition
moment the model yields
\beq
\mu(\Delta^{++})={3\over 2}(g_8+3g_3)=5.55\, n.m.\quad
\mu(\Delta\rightarrow N)=2\sqrt{2}g_3 = 2.65\, n.m.\; ,
\eeq
which may be compared to the corresponding empirical
values $4.52$ n.m. \cite{Boss} and $3.1$ n.m. \cite{PRPD}
respectively.

The nucleon elastic form factors are  then
\beqarray
G_E^p(Q^2)&=&\half[f_3+f_8] 
F_0(\eta), \quad G_E^n(Q^2)=\half [f_3-f_8]\, F_0(\eta)\; ,
\cr\cr
G_M^p(Q^2)&=&\mu_p F_0(\eta), \quad G_M^n(Q^2)=\mu_n F_0(\eta)\; ,
\eeqarray
with $Q^2=4m_p^2\eta$. The 
magnetic form factors of the nucleons
obtained with the quark mass (5.9) 
compare well with the empirical parametrization
\cite{Hoe}. Observed features could be reproduced to any accuracy
by assuming a suitable $\eta$ dependence for
the quark form factors $f_3$, $f_8$, $g_3$ and $g_8$.

Equivalent results may obviously be obtained with other 
forms of kinematics. For instance
\beq
m^2(\tilde Q^2,\eta)= \omega^2{1+\eta\over 3\eta}\,ln(1+\tilde Q^2/.71)\; ,
\label{INST}
\eeq
yields again the dipole form for the elastic form factors, but
differences will appear in the relations to transition amplitudes.

\newpage

\sect{Transition Form Factors}
\setcounter{equation}{0}

Current conservation implies 
that the transition matrix elements must satisfy
\beq
(M'v'-Mv)\cdot I = \half (M'+M)(v'-v)\cdot I +(M'-M)v^0  u\cdot I=0\; .
\eeq
Since $(v'- v)\cdot I$ vanishes the second term must be cancelled by
an appropriate interaction current. Non-vanishing longitudinal form factors 
depend  on model dependent interaction currents.

All transition form factors are functions of
$\eta$ multiplied by spin-flavor structure matrix elements.
For each transition the dependence on $Q^2$ is given by the general relation
$
Q^2= 4MM'\eta -(M'-M)^2\; ,
$
which implies
$
\;\eta =\eta_{rad}:=(M'-M)^2/ 4 M'M\; 
$
for real-photon transitions.

The  helicity amplitudes (\ref{HEL}) are products
of functions, which depend only on the spatial wave functions 
multiplied
by spin-flavor amplitudes:
\beq
A_\lambda(Q^2)
=\sqrt{4\pi\alpha}\,\sqrt{{2M'\eta\over {M'}^2-M^2}} 
F_{N,L}(\eta)\; \A_\lambda(T,S,j)\; .
\eeq
Here $F_{0,0}(\eta)\equiv F_0(\eta)$ and
\beq
F_{2,2}(\eta)=\sqrt{2}F_{2,0}(\eta)=
\sqrt{2\over 3}\,{6m^2(\eta)\eta\over \omega^2( 1+\eta)}F_0(\eta)\; ,
\qquad
F_{1,1}(\eta):=
{\sqrt{6}m(\eta)\over \omega} \,
\sqrt{{\eta\over 1+\eta}}F_0(\eta)\; ,
\eeq
The spin-isospin factors,
\beq
\A_\lambda(T,S,j):=(L,S,0,\lambda|j,\lambda)\, \nhalf\bra{\tau,\lambda }
\Phi_{S,T}^\dagger[\turd g_8 +\tau^{(3)}_z\,g_3]
i\sigma_y^{(3)}
\Phi_{\half,\half}\ket {\lambda-1,\tau}\; ,
\eeq
for transitions to the states in Table 2 are tabulated in Table 3
(the spin-flavor matrix elements depend indirectly on the
spatial wave function by the requirement that the 
baryon states be symmetric).
The helicity amplitudes obtained  with these expressions are
compared to the corresponding empirical ones given
in ref. \cite{PRPD} in Tables 4 and 5. The model  helicity
amplitudes were calculated using the  model
mass values in Table 1 in the kinematic 
expressions. The magnitudes of the model  helicity
amplitudes 
for photon decay are
similar to those of the harmonic oscillator quark model
\cite{Cop}, but the decrease with increasing $Q^2$ values
is much slower. Related to this is the fact that in the present 
point-form
impulse approximation  the $N\rightarrow \Delta(1232)$
magnetic transition form factor falls off at a slower
rate with $Q^2$ than the dipole form factor, in disagreement
 with present data \cite{Bur}.  This disagreement
is a robust feature of the point-form impulse approximation
not shared by other forms of kinematics. For instance, 
$N\rightarrow \Delta(1232)$ transition amplitudes
obtained with (\ref{INST}) decrease faster than the corresponding
nucleon form factor. The choice of the kinematics determines
the relative role of impulse  and interaction currents.
Reduction of the present uncertainty 
 of the
empirical helicity amplitudes should 
provide a more definite indication of the preferred kinematics.

The present results share the qualitative feature
of other quark  models that the  magnitudes of most helicity 
amplitudes are reasonable, the main exception being the
spin-3/2 negative parity multiplet, where configuration
mixing is needed for better overlap with the empirical
values \cite{Bijk}. 

\newpage

\sect{Summary}
We have outlined a Poincar\' e
covariant approach to  electromagnetic 
form factors and transition amplitudes
of baryons.
The present, deliberately oversimplified, model was
designed to elucidate the qualitative differences between
baryon structure as described in terms of constituent
quark and the structure of the few nucleon systems.
Exact realization of \Poincare covariance is essential
for the former. A 
few-nucleon system is essentially a system of free
nucleons with a binding correction added to the free
particle mass operator, whereas
a system of confined quarks is  
described by a degenerate
mass operator, with modifications 
that yield the required empirical mass splittings.
In the description of confined quark the \Poincare represtenation
with the degenerate mass operator plays a role similar to that 
of the free particle representation in the description of 
few-nucleon systems.

The point-form kinematics provides
a particularly simple framework for the description
of transition observables, when the momentum transfer
ranges over both space- and time-like values.

A key element in the approach described above
is a confining mass
operator with a spin- and flavor dependent hyperfine
term, the eigenfunctions of which are symmetrized 
products of orbital wave functions and spin-flavor
functions. This mass operator
provides a satisfactory account of the
empirical spectra of the non-strange baryons, and may with
minor adjustments be applied to the spectra of the
strange \cite{GlRi} and heavy flavor hyperons as
well \cite{GloRis}. 
The electromagnetic current model was constructed to
implement the same qualitative features.
The current matrix elements are integrals, which involve
only the orbital wave functions multiplied by the
spin-flavor matrix elements. The impulse approximation
provides definite relations of transition amplitude to
ground-state properties depending on the form of kinematics.
It should be emphasized that the framework  leaves
considerable freedom in the construction of  current
models. 
A more elaborate 
version would involve 
dependence of the kernels (\ref{LEAN}) on 
quark velocities as well as  the
spin and flavor variables \cite{Coe1}. Such dependence
is required for the inclusion of a convection current,
as well as for a realistic description of the axial
vector structure of the baryons.

	Additional features that are readily incorporated
are a non-vanishing neutron charge form factor 
\cite{Ko} -- e.g. by
including $SU(3)_F$ breaking quark form factors in the
charge operator in (5.5) -- and a tensor component in the
hyperfine term in the mass operator. A tensor component
is required for a non-vanishing E2/M1 ratio for the
$\Delta(1232)\rightarrow N$ decay. Two-- and three--quark
current operators may of course also be added to the
model, but there is no obvious need for such currents.
The definition of single-quark currents 
is, of course, strongly model dependent.

\vspace{1cm}

\centerline{{\bf Acknowledgments}}
We thank T.-S. H. Lee, W. H. Klink and W. N. Polyzou for
many instructive discussions on this topic. This work was 
supported in part by the Department of
Energy, Nuclear Physics Division, under contracts 
W-31-109-ENG-38 and by the Academy of Finland under contract
34081.

\vspace{1cm}

\newpage
\centerline{\bf Table 1}
\vspace{0.5cm}

The nucleon and $\Delta$-states up to $\sim$ 1700 MeV. The column
$\epsilon$ contains the eigenvalues of the mass operator (3.4).
The average over the multiplet of the real part of the
empirical pole positions is denoted EXP. The values predicted
by the mass operator (3.4) are listed (in brackets) 
below the empirical values.

\vspace{1cm}

\vspace{1cm}

\begin{center}
\begin{tabular}{l|l|l|l} \hline
$NL[f]_{FS}[f]_F[f]_S$ & $LS$ Multiplet & EXP & $\epsilon$
\\ 
 & & (model value) & \\ \hline
$00[3]_{FS}[21]_F[21]_S$ & ${1\over 2}^+, N$ & 939 &
$\sqrt{18}\omega-14C$\\
 & & (940) & \\
$00[3]_{FS}[3]_F[3]_S$ & ${3\over 2}^+,\Delta$ & 1211 &
$\sqrt{18}\omega-4C$\\
 & & (1211) & \\
$20[3]_{FS}[21]_F[21]_S$ & ${1\over 2}^+,N(1440)$ & 1346 &
$\sqrt{30}\omega-14C$\\
 & & (1324) & \\
$11[21]_{FS}[21]_F[21]_S$ & ${1\over 2}^-,N(1535),{3\over
2}^-N(1520)$ & 1508 & $\sqrt{24}\omega-2C$\\
 & & (1554) & $+2A$\\
$20[3]_{FS}[3]_F[3]_S$ & ${3\over 2}^+,\Delta(1600)$ & 1675 &
$\sqrt{30}\omega-4C$\\
 & & (1595) & \\
$11[21]_{FS}[3]_F[21]_S$ & ${1\over 2}^-,\Delta(1620);{3\over
2}^-,\Delta(1700)$ & 1620 & $\sqrt{24}\omega+4C$\\
 & & (1718) & $+2A$\\
$11[21]_{FS}[21]_F[3]_S$ & ${1\over 2}^-,N(1650);{3\over
2}^-,N(1700)$ & 1679 & $\sqrt{24}\omega+2C$\\
 & ${5\over 2}^-,N(1675)$ & (1664) & $+2A$\\
$2(20)2[3 ]_{FS}[21]_F[21]_S$ & ${3\over 2}^+,N(1720),{5\over
2}^+,N(1680)$ & 1693 & $\sqrt{30}\omega-14C$\\
 & & (1582) & $+6A$\\ \hline
\end{tabular}
\end{center}

\newpage
\centerline{\bf Table 2}
\vspace{0.5cm}

Explicit wave functions for the baryon states in Table 2. The
functions $\varphi_{nlm}$ are harmonic oscillator wave functions.
The total
angular momentum is denoted J and the 3rd
components of the spin and isospin are denoted $S_3$ and $T_3$
respectively. The subscripts $\pm$ on the spin-isospin states 
indicate are shorthands for the Yamanouchi symbols (112) and
(121) respectively.

\vspace{1cm}

\begin{center}
\begin{tabular}{|l|l|} \hline
&\\
$p,n,{1\over 2}^+$ & ${1\over \sqrt{2}}\varphi_{000}(\vec \kappa)\,
\varphi_{000}(\vec k)
\left\{|{1\over 2},T_3\>_{_+}|{1\over 2},S_3\>_{_+}+|{1\over
2},T_3\>_{_-}|{1\over 2},S_3\>_{_-}\right\}$\\ 
&\\ \hline
&\\
$\Delta(1232),{3\over 2}^+$ & $\varphi_{000}(\vec
p)\,\varphi_{000}(\vec k)|{3\over 2},T_3\>|{3\over 2},S_3\>$\\
& \\ \hline
$N(1440),{1\over 2}^+$ & ${1\over 2}\,\left\{
\varphi_{200}(\vec \kappa)\,
\varphi_{000}(\vec k)+\varphi_{000}(\vec \kappa)\,\varphi_{200}(\vec k)
\right\}$ \\ 
&\\
&$\left\{|{1\over 2},T_3\>_{_+}|{1\over 2},S_3\>_{_+}+|{1\over
2},T_3\>_{_-}|{1\over 2},S_3\>_{_-}\right\}$\\
 & \\ \hline
$N(1535),{1\over 2}^-$ & ${1\over 2}\sum_{m s}(1,{1\over2,}m,s|J,S_3)
\Bigl\{\varphi_{01m}(\vec \kappa)\,\varphi_{000}(\vec k)$\\
$N(1520),{3\over 2}^-$ & \\
 & $\left[|{1\over 2},T_3\>_{_+}|{1\over 2},s\>_{_+}-|{1\over 2},T_3\>_{_-}
|{1\over 2},s\>_{_-}\right]$\\
 & \\
 & $+\varphi_{000}(\vec \kappa)\,\varphi_{01m}(\vec k)\left[|{1\over
2},T_3\>_{_+}|{1\over 2},s\>_{_-}
+|{1\over 2},T_3\>_{_-}|{1\over 2},s\>_{_+}\right]\Bigr\}$\\ \hline
& \\
$\Delta(1600)$ & ${1\over \sqrt{2}}\Bigl\{\varphi_{200}(\vec \kappa)
\,\varphi_{000}
(\vec k)+\varphi_{000}(\vec \kappa)\,\varphi_{200}(\vec k)\Bigr\}
|{3\over 2},T_3\>|{3\over
2}S_3\>$\\
 & \\ \hline
$\Delta(1620),{1\over 2}^-$ & ${1\over \sqrt{2}}\sum_{ms}
(1,{1\over 2},m,s|J,S_3)
\Bigl\{\varphi_{01m}\,(\vec \kappa)\varphi_{000}(\vec k)|{3\over
2},T_3\>|{1\over 2},s\>_{_+}$\\
$\Delta(1700),{3\over 2}^-$ & \\
 & $+\varphi_{000}(\vec \kappa)\varphi_{01m}(\vec k)|{3\over 2},T_3\>
|{1\over 2},s\>_{_-}\Bigr\}$\\ \hline
$N(1650),{1\over 2}^-$ & ${1\over \sqrt{2}}\sum_{ms}(1,{3\over 2},m,s|J,S_3)
\Bigl\{\varphi_{01m}
(\vec \kappa)\,\varphi_{000}(\vec k)|{1\over2},T_3\>_{_+}$\\
$N(1700),{3\over 2}^-$ & \\
$N(1675),{5\over 2}^-$ & $+\varphi_{000}(\vec \kappa)\,\varphi_{01m}(\vec
q)|{1\over 2},T_3\>_{_-}\Bigr\}|{3\over 2},s\>$\\ \hline
$N(1720),{3\over 2}^+$ & ${1\over 2}\sum_{ms}(2,{1\over
2},m,s|JS_3)\Bigl\{\varphi_{22m}(\vec \kappa)\varphi_{000}(\vec k)
+\varphi_{000}(\vec \kappa)\varphi_{22m}(\vec k)\Bigr\}$\\
$N(1680),{5\over 2}^+$ & $\Bigl\{|{1\over 2},T_3\>_+|{1\over
2},s\>_++|{1\over 2},T_3\>_-|{1\over 2},s\>_-\Bigr\}$\\ \hline
\end{tabular}
\end{center}

\newpage
\centerline{\bf Table 3}
\vspace{0.5cm}

The spin-isospin factors (6.4) for the helicity amplitudes for 
$p\to N^*$ and $p\to \Delta$
transitions. For transitions from the neutron to nucleon resonances
the sign of the
terms containing $g_3$ should be reversed.
 
\vspace{1cm}

\begin{center}
\begin{tabular}{|l|l|l|} \hline
&& \\
& $\qquad\A_\half$&$\qquad\A_\thalf$  \\
&& \\ \hline
$\Delta(1232),{3\over 2}^+$ & 
$\qquad -\sqrt{2}\,g_3$ &$\qquad -\sqrt{6}\,g_3\qquad$\\
\hline
$N(1440),{1\over 2}^+$ & $\qquad\half\left[g_8+5g_3\right]\;$ &\\
\hline
$N(1535),{1\over 2}^-$ & $\qquad -\sqrt{1\over12}
\left[g_8+2g_3\right] $ &
\\
& &\\
$N(1520),{3\over 2}^-$ & $\qquad { \sqrt{1\over 6}}\;\left[g_8+
2g_3\right]
\qquad$& $\;\qquad 0.0$                      \\
\hline
$\Delta(1600),{3\over 2}^+$ & $\qquad -\sqrt{2}\,g_3
$&$\qquad -\sqrt{6}\,g_3$\\
\hline
$\Delta(1620),{1\over 2}^-$ & $\qquad-\sqrt{1\over 3}\, g_3\quad $&\\
& &\\
$\Delta(1700),{3\over 2}^-$ & $\qquad +\sqrt{2\over 3}\, 
g_3$&$\qquad 0.0$ \\
\hline
$N(1650),{1\over 2}^-$ & $\quad -{1\over 2\sqrt{3}}\;[g_8-g_3]\quad$&\\
& &\\
$N(1700),{3\over 2}^-$ &$\qquad -{1\over 2\sqrt{15}}\;[g_8-g_3]$&
 $\quad -{3\over 2\sqrt{5}}[g_8-g_3]\quad$\\
& &\\
$N(1675),{5\over 2}^-$ & $\qquad {1\over 2}\sqrt{3\over 5}
\;[g_8-g_3]$&$\quad \sqrt{3\over 10}[g_8-g_3]\quad$ \\
 \hline
$N(1720),{3\over 2}^+$ & $\qquad-{1\over \sqrt{10}}\;
\left[g_8+5g_3\right] $& $\qquad 0.0$\\
& &\\
$N(1680),{5\over 2}^+$ & $\qquad \sqrt{3\over 20} \;\left[g_8+
5g_3\right] $
 &$\qquad 0.0$\\ 
\hline
\end{tabular}
\end{center}
\newpage
	
\newpage
\centerline{\bf Table 4}
\vspace{0.5cm}
Comparison of the model helicity amplitudes for
nucleon resonance decays (in units of GeV$^{-\half}$) 
to the corresponding empirical
values (Data) from ref. \cite{PRPD}.
\vspace{1cm}
\begin{center}
\begin{tabular}{|l|l|l|l|l|} \hline
& & & &\\
 &  $\quad N^*\to p \gamma\; $ & $\quad N^*\to n \gamma\; $&
$\qquad N^* \to p\gamma\qquad$ & $\qquad N^* \to n\gamma\qquad$ \cr
& $\qquad$ Data & $\quad$ Data& $\qquad$ Model & $\qquad$ Model
\cr & & & &\\ \hline
$N(1440)$ &&&&\\
$A_{1/2}$ & $\quad -0.065\pm0.004
$&$\quad+0.040\pm 0.010\quad$ & $\quad +0.022\quad$ & $\quad -0.014$\\ \hline
& & & &\cr
$N(1535)$ &&&&\\
$A_{1/2}$ &$\quad +0.070\pm 0.012 $ 
& $\quad -0.046\pm 0.027 \quad$  & $\quad +0.036$ & $\quad-0.013$\\ \hline
$N(1520)$ &&&&\\
$A_{1/2}$ & $\quad -0.024\pm 0.009$ & 
$\quad -0.059\pm 0.009$ & $\quad -0.051$ & $\quad +0.018$
\cr $A_{3/2}$ &$\quad +0.166\pm0.005$ & $\quad -0.139\pm0.011$ &
$\quad +0.0$ & $\quad +0.0$\\ \hline
$N(1650)$ &&&&\\
$A_{1/2}$ &$\quad +0.053\pm0.016$ &
$\quad -0.015\pm0.021$&$\quad +0.001$ & $\quad -0.024$\\ \hline
$N(1700)$ &&&&\\
$A_{1/2}$ &$\quad -0.018\pm0.013$ &
$\quad +0.001\pm0.050$&$\quad +0.001$ & $\quad +0.092$\cr
$A_{3/2}$ & $\quad -0.002\pm0.024$ & $\quad -0.003\pm0.044$ &
$\quad -0.002$ & $\quad +0.048$\\ \hline
$N(1675)$ &&&&\\
$A_{1/2}$ &$\quad +0.019\pm0.008$ &
$\quad -0.043\pm0.012$&$\quad +0.001$ & $\quad -0.034$\cr
$A_{3/2}$&$\quad +0.015\pm0.009$ &$\quad -0.058\pm0.013$&
$\quad -0.002$ & $\quad +0.048$\\ \hline
$N(1720)$ &&&&\\
$A_{1/2}$&$\quad +0.018\pm0.030 $&$\quad +0.001\pm0.015$&
$\quad -0.052$ & $\quad +0.035$\cr
$A_{3/2}$&$\quad -0.019\pm0.020$&$\quad -0.029\pm0.061$&
$\quad +0.0$ & $\quad +0.0$\\ \hline
$N(1680)$ &&&&\\
$A_{1/2}$ &$\quad -0.015\pm0.006 $&$\quad +0.029\pm0.010$
&$\quad +0.064$  & $\quad -0.043$\cr
$A_{3/2}$&$\quad +0.133\pm0.012$&$\quad -0.033\pm0.009$
&$\quad +0.0$ & $\quad -0.0$\\ \hline
\end{tabular}
\end{center}
\newpage
\centerline{\bf Table 5}
\vspace{0.5cm}
Comparison of the model helicity amplitudes for
$\Delta$ resonance decays (in units of GeV$^{-\half}$)
to the corresponding empirical
values (Data) from ref. \cite{PRPD}
\vspace{1cm}

\begin{center}
\begin{tabular}{|l|l|l|} \hline
& &\\
 &  $\quad \;\quad \Delta \to N\gamma$ & 
$\qquad\Delta \to N \gamma\qquad$ \cr & $\qquad$ Data &$\qquad$ Model
\\ \hline
$\Delta(1232)$ &&\\ 
$A_{1/2}$ &$\quad - 0.140\pm 0.005$ & $\;\; -0.089$\cr
$A_{3/2}$& $\quad   -0.258 \pm 0.006$ & $\;\;-0.15$\\ \hline
$\Delta(1600)$ &&\\
$A_{1/2}$ &$\quad -0.023\pm 0.020$ 
& $\;\; -0.020$  \\
$A_{3/2}$ &$ \quad -0.009\pm 0.021$ & $\;\; -0.034$\\ \hline
$\Delta(1620)$ &&\\
$A_{1/2}$ &$\quad +0.027\pm 0.011$ & $\;
\;+0.038 $\\ \hline
$\Delta(1700)$ &&\\
$A_{1/2}$ & $\quad +0.104\pm 0.015$ &$\;\;
-0.054 $
\cr $A_{3/2}$&  $ \quad +0.085\pm 0.022$&$\;\;+0.0 $\\ \hline
\end{tabular}
\end{center}

\vspace{1cm}


\begin{thebibliography}{99}
\bibitem{Haa}For a concise overview see R. Haag, {\it Local Quantum Physics}
Sec. I.3.2, Springer-Verlag (1993)
\bibitem{Kei} B.D. Keister and 
W. N. Polyzou, Advances  in Nuclear Physics {\bf 20}
\bibitem{Bijk} R. Bijker, F. Iachello and A. Leviatan, Ann. Phys.
{\bf 236}, 69 (1994)
\bibitem{Fu} K. Fujimura, T. Kobayashi and M. Namiki, Prog. Theory
Phys. {\bf 43} (1970) 73, {\bf 44} (1970) 193
\bibitem{Rav1} R. P. Feynman, M. Kislinger and F. Ravndal,
Phys. Rev. {\bf D3}, 2706 (1971)
\bibitem{Dir} P. A. M. Dirac, Reviews of Modern Physics 
{\bf 21}, 392 (1949)
\bibitem{BeTe} V. B. Beretetskii and M. V. Terentev, Sov. J.
Nucl. Phys. {\bf 24}, 547 (1976)
\bibitem{ChCo} P. L. Chung and F. Coester, Phys. Rev. {\bf D44},
229 (1991)
\bibitem{CaKe} S. Capstick and B. D. Keister, Phys. Rev. {\bf D51},
3598 (1995)
\bibitem{Carda} F. Cardarelli et al., Phys. Lett. {\bf B357}, 267
(1995)
\bibitem{Cardb} F. Cardarelli et al., Phys. Lett. {\bf B371}, 7 (1996)
\bibitem{Rav2} F. Ravndal, Phys. Rev. {\bf D4}, 1466 (1971)
\bibitem{Jall} H. Jallouli and H. Sadzjian, Ann. Phys.
{\bf 253}, 376 (1997)
\bibitem{Tjon} M. J. Zuilhof and J. A. Tjon, Phys. Rev. {\bf C22},
2369 (1980)
\bibitem{Rob} C. D. Roberts and A. G. Williams, Prog. Part. 
Nucl. Phys. {\bf 33}, 477 (1994); P. C. Tandy, Prog. Part. 
Nucl. Phys. {\bf 39} (1997), e-print archive nucl-th/9705018
\bibitem{Ost} K. Osterwalder and R. Schrader, Comm. Math. Phys.
{\bf 31}, 83 (1973); {\bf 41}, 281 (1975); J. Glimm and
A. Jaffe, {\it Quantum Physics}, Springer-Verlag, New York (1987)
\bibitem{GlRi} L. Ya. Glozman and D. O. Riska, Phys. Rep.
{\bf 268}, 263 (1996)
\bibitem{Hoe} G. H\"{o}hler et al. Nucl. Phys. {\bf B114}, 505 (1976)
\bibitem{Boss} A. Bosshard et al., Phys. Rev. {\bf D44} 1962  (1991)
\bibitem{PRPD}Particle Data Tables Phys. Rev. {\bf D 54} 601 (1996)
\bibitem{Cop} L. A. Copley, G. Karl and E. Obryk, 
Nucl. Phys. {\bf B13}, 303 (1969)
\bibitem{Bur} V. D. Burkert, Int. J. Mod. Phys. {\bf E1}, 421 (1992)
\bibitem{GloRis} L. Ya. Glozman and D. O. Riska, Nucl. Phys.
{\bf A603}, 326 (1996)
\bibitem{Coe1} F. Coester and D. O. Riska, in preparation
\bibitem{Ko} L. Koester, W. Nistler and W. Waschkowski, Phys. Rev.
Lett. {\bf 36}, 1021 (1976)
\end{thebibliography}
\end{document}